# Phonon Anomaly In 9R-Ba$_{0.9}$Sr$_{0.1}$MnO$_3$

Bommareddy Poojitha[1] and Surajit Saha[1, a)]

[1]Indian Institute of Science Education and Research, Bhopal, Bhauri campus 462066

*a) surajit@iiserb.ac.in*

**Abstract.** Materials which possess coupling between different degrees of freedom such as charge, spin, orbital and lattice are of tremendous interest due to their potential for device applications. It has been recently realized that some of manganite perovskite oxides show structural and ferroelectric instabilities which may be tuned by imposing internal or external strain. Among manganite perovskites, AMnO$_3$ (A = Ca, Sr and Ba) are reported to be good candidates to show spin-phonon coupling driven multiferroicity. Here, we have probed the correlation between vibrational and magnetic properties in the 9R-type hexagonal Ba$_{0.9}$Sr$_{0.1}$MnO$_3$ (BSM10). Nine Raman active modes are observed which are characteristic phonon modes for 9R phase. The frequency ($\omega$) of an E$_g$ (344 cm$^{-1}$) mode which is related to Mn vibrations shows anomaly with temperature (T) manifested as a change in slope of $\omega$ vs T at 260 K. The temperature dependent magnetization of BSM10 suggests a possible antiferromagnetic ordering at ~ 260 K. We, therefore, attribute the phonon anomaly to spin phonon coupling arising due to the magnetic ordering.

## INTRODUCTION

Perovskite oxides show multifunctional properties due to the presence of couplings among various degrees of freedom such as charge, spin, orbital, and lattice. Especially, AMnO$_3$ with A = divalent cation compounds are interesting to study because of their richness in structural phases which can be a platform to play with the coupling between different degrees of freedom [1]. Structure of these compounds can be rationalized by calculating *Goldschmidt* tolerance factor *t*, ($t = \frac{<A-O>}{\sqrt{2}<Mn-O>}$) [2]. *t* = 1 gives rise to ideal cubic structure (e.g. SrTiO$_3$). In cubic close packing (ccp) structure, all MnO$_6$ octahedra are corner shared. For $t < 1$, reduction in the Mn-O-Mn bond angle due to smaller A cation introduces distortion in MnO$_6$ octahedra which leads to distorted cubic structure (e.g. CaMnO$_3$). On the other hand, *t* > 1 give rise to hexagonal structure in which MnO$_6$ octahedra will be linked by face-sharing in order to provide space for larger A cation. Type of hexagonal structure vary depends on epex-sharing connectivity. In 2H-type structure, hexagonal close packing (hcp) of AO$_3$ layers leads to infinite strings of face-shared MnO$_6$ octahedra along c-axis. Different ratios of ccp and hcp arrangements may lead perovskite manganites to adapt different types of hexagonal structures [3].

Among all, BaMnO$_3$ has drawn an enormous attention due to its polymorphic nature which is tunable by growth conditions and/or chemical doping [4]. It can be stabilized in either of 2H, 4H, 6H, 8H, 10H, 9R, 12R, and 15R type hexagonal structure at room temperature. In this notation, integer represents the number of layers, H is hexagonal symmetry and R is rhombohedral symmetry in the unit cell. For example, 2H phase has two layers in the unit cell with "*ab*" stacking sequence. Similarly, nine layers per unit cell with the stacking sequence "*ababcbcac*" exists for the 9R phase. SrMnO$_3$ is antiferromagnetic insulator which can adapt either cubic or 4H phase with corner-sharing Mn$_2$O$_9$ dimers. However, hexagonal SrMnO$_3$ has higher Neel temperature (~ 280 K) than in its cubic phase (~ 230 K). Coupling between vibrational and magnetic properties is expected in Ba doped SrMnO$_3$ due to off centering of Mn$^{+4}$ ions [5]. Here, we report thermal expansion, vibrational and magnetic properties of 9R-Ba$_{0.9}$Sr$_{0.1}$MnO$_3$ (9R-BSM10). The parameters for volumetric thermal expansion coefficient are calculated. All characteristic Raman active modes for 9R phase are observed which are in agreement with group theory calculation. The frequency of E$_g$ (344 cm$^{-1}$) mode is found to be anomalous with temperature and shows the slope change in $\omega$ vs T around 260 K which can be attributed to spin phonon coupling originated from antiferromagnetic ordering in the sample.

## SAMPLE PREPARATION AND EXPERIMENTAL TECHNIQUES

The polycrystalline $Ba_{0.9}Sr_{0.1}MnO_3$ sample is prepared using a conventional solid-state reaction method. $BaCO_3$, $SrCO_3$, and $MnO_2$ in a stoichiometric ratio are taken as precursors. The well mixed powder is calcinated at 800°C, 900°C and 1000°C for 24 hours each time with intermediate grinding. Final sintering is done at 1200°C for 12 hours [4]. X-ray diffraction measurements are carried out using PANalytical Empyrean x-ray diffractometer with Cu-$K_\alpha$ radiation of wavelength 1.5406 nm. Anton paar TTK 450 liquid $N_2$ based heating system is used for temperature dependent measurements. Inelastic light scattering measurements are done in backscattering configuration using LabRAM HR Evolution Horiba Scientific Raman spectrometer with 532 nm laser as an excitation line. Linkam stage of type HFS600E-PB4 is used for temperature dependent Raman measurements. DC Magnetization measurements were carried out using Quantum Design SQUID-VSM (Superconducting Quantum Interference Device with Vibrating Sample Magnetometer).

## RESULTS AND DISCUSSION

*Raman spectroscopy:* Fig. 1a shows the Raman spectrum of $Ba_{0.9}Sr_{0.1}MnO_3$ at room temperature. 9R phase has space group R-3m (No. 166) and point group $D_{3d}$ having three molecules per unit cell. The zone center Raman active modes predicted by group theory for 9R-BSM10 can be represented by $\Gamma_{Raman} = 4A_{1g}+5E_g$ [6]. The Raman spectrum is analyzed with Lorentzian function to find the frequencies and corresponding line widths of phonon modes. Nine peaks are observed as expected from group theory. The frequency of modes falls at about 100 [$E_g$], 255 [$A_{1g}$], 344 [$E_g$], 408 [$A_{1g}$], 424 [$E_g$], 531 [$E_g$], 563 [$A_{1g}$], and 637 [$A_{1g}$] cm$^{-1}$. Corresponding symmetries are given in square brackets which are assigned based on the report on 9R- $BaRuO_3$ [6]. Low-frequency modes involve vibrations of Ba/Sr and Mn atoms while the high frequency modes are due to oxygen vibrations. The frequencies of the modes at 255 cm$^{-1}$ and 344 cm$^{-1}$ modes as a function of temperature are shown in Fig. 1 (b, c). Frequency of 255 cm$^{-1}$ mode decreases with increasing temperature which can be attributed to thermal expansion of the lattice due to quasi-harmonic behaviour. Change in phonon frequency with temperature due to cubic anharmonicity can be expressed as, [7]

$$\omega_{anh}(T) = \omega_0 + A\left[1 + \frac{2}{(e^{\frac{\hbar\omega_0}{2k_BT}}-1)}\right] \quad (1)$$

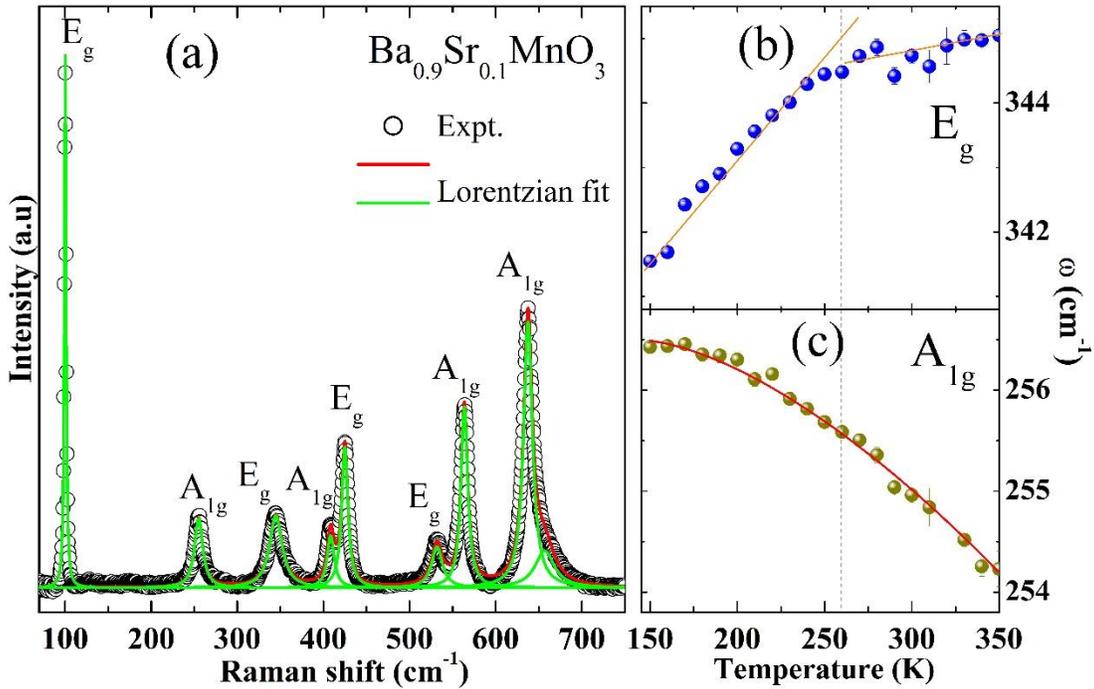

**FIGURE 1.** *(a) Raman spectrum of polycrystalline $Ba_{0.9}Sr_{0.1}MnO_3$ collected with excitation wavelength 532 nm at room temperature. Frequency (ω) of $E_g$ modes associated with (b) Mn and (c) Ba/Sr vibrations as a function of temperature. Solid lines are drawn for guide to eye.*

Where, $\omega_0$ is the frequency phonon at absolute zero temperature, A is cubic anharmonic coefficient for frequency, $\hbar$ is reduced plank constant, $k_B$ is Boltzmann constant, and T is variable temperature. Cubic anharmonic coefficients are extracted by fitting temperature dependent phonon frequency with Eq. 1. The $E_g$ mode at 344 cm$^{-1}$ show anomalous trend with temperature i.e. increase in frequency with increasing temperature, and shows a change in slope at 260 K which cannot be explained by cubic phonon anharmonicity (Eq. 1). In order to verify the association of any structural changes in BSM10 at 260 K to the change in slope in $E_g$ (344 cm$^{-1}$) mode, we have performed temperature dependent XRD which is discussed below.

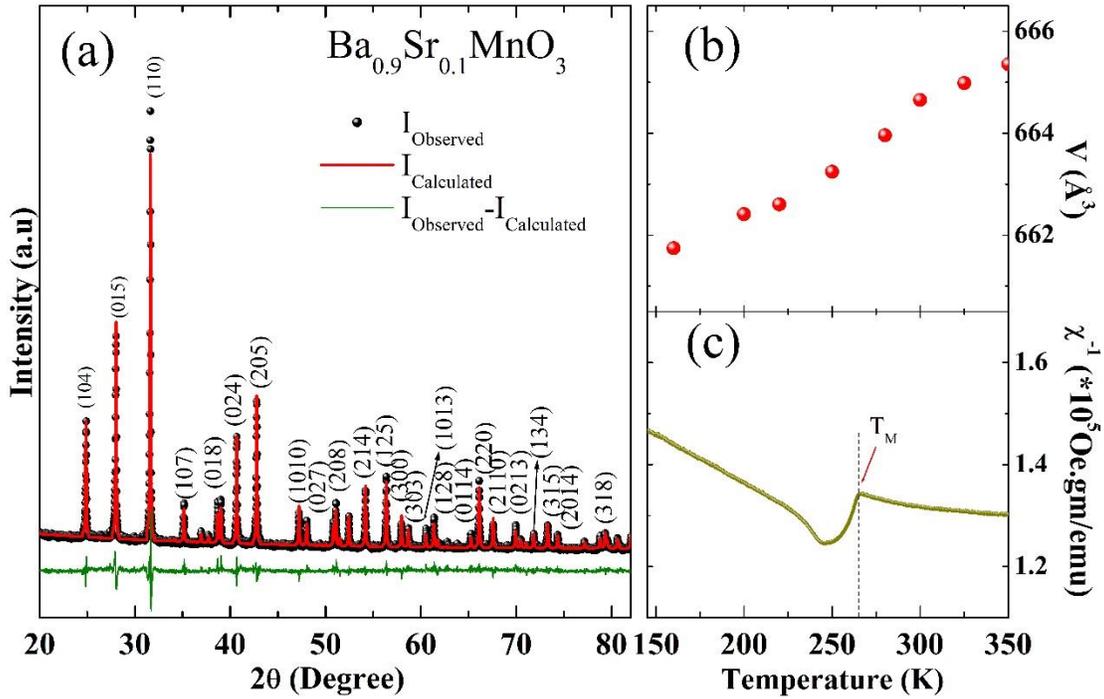

**FIGURE 2**. *(a) X-ray diffraction pattern for the polycrystalline $Ba_{0.9}Sr_{0.1}MnO_3$ at room temperature, (b) unit cell volume as a function of temperature. (c) Inverse magnetic susceptibility as a function of temperature measured with applied dc field of 500 Oe. $T_M$ is the magnetic transition.*

*X-ray diffraction:* X-ray diffraction data at room temperature is refined with the Rietveld method using High Score Plus software. It shows that the sample is stabilized in 9R-type hexagonal structure with the space group R-3mH (No.166) (Fig.2a). The refined lattice parameters are a = b = 5.6445 Å and c = 20.8969 Å at room temperature. The obtained XRD reflections at various temperatures and corresponding (*hkl*) values reveal that the structure of BSM10 remains in 9R-type hexagonal down to lowest measured temperature in our experiments. Unit cell volume is plotted against temperature as shown in fig. 2b. The unit cell volume as a function of temperature can be expressed as [8-10], $V(T) = V_{T_r} \exp[\int_{T_r}^{T} \alpha(T)dT]$. Where $T_r$ is reference temperature (here it is 150K). $V_{T_r}$ is volume at $T_r$, $\alpha(T)$ is volumetric thermal expansion coefficient having the form $\alpha(T) = a_0 + a_1T$, value of $\alpha(T)$ at each temperature can be calculated from experimental data (V vs T) using the expression $\alpha(T) = \frac{1}{V}\frac{dV}{dT} = \frac{1}{V_{LT}}\frac{V-V_{LT}}{T-T_{LT}}$. Where, $V_{LT}$ is volume at lowest measured temperature. The parameters extracted from above expressions ($a_0$, $a_1$) for volumetric thermal expansion coefficient are $0.2 \times 10^{-4}$ K$^{-1}$ and $1.9145 \times 10^{-8}$ K$^{-2}$ respectively. Detailed analysis of temperature dependent XRD reveals that, there is no structural transition or anomaly in thermal expansion of the material in the measured temperature range. In order to probe any possible involvement of magnetic ordering in the phonon anomaly (E$_g$ mode at 344 cm$^{-1}$), we have performed temperature dependent magnetization measurements.

*Magnetism:* Fig. 2c shows the inverse magnetic susceptibility plotted as a function of temperature. The inverse susceptibility appears to be weakly temperature dependent below 350 K down to 260 K. However, below 260 K, $\chi^{-1}$ shows an abrupt increase with decreasing temperature thus indicating a possible antiferromagnetic ordering below 260 K. This observation is similar to the reported antiferromagnetism in BaMnO$_{3-\delta}$ [11] and Sr$_{0.6}$Ba$_{0.4}$MnO$_3$ [12].

*Spin-phonon coupling:* The phonon anomaly and change in slope at T$_M$ ~ 260 K is evident in 344 cm$^{-1}$ (E$_g$) mode, shown in fig. 1(b), which is associated with Mn vibrations. Whereas, 255 cm$^{-1}$ (E$_g$) mode involving Ba/Sr vibrations doesn't show any anomaly or response to the magnetic transition temperature. The observed anomaly in E$_g$ (344 cm$^{-1}$) mode with the temperature and the change in slope at T$_M$, may therefore be associated with the antiferromagnetic ordering observed below T$_M$ (as shown in fig. 2c). The antiferromagnetic ordering of Mn ions which are also involved in the 344 cm$^{-1}$ (E$_g$) phonon mode may give rise to spin phonon coupling [12].

# CONCLUSION

Magnetic ordering and spin phonon interactions are of great interest in perovskite manganites. Here, we synthesized 9R phase of Ba$_{0.9}$Sr$_{0.1}$MnO$_3$ using solid state reaction route. Raman spectra of BSM10 show all the characteristic modes of 9R phase. Temperature dependent Raman spectra show an anomalous behaviour of the Mn atomic vibrations at 344 cm$^{-1}$ (E$_g$) with a change in slope at 260 K. Temperature dependent XRD rules out any possibility of structural phase transition. However, signature of antiferromagnetism has been identified in the magnetization data. Therefore, the phonon anomaly is attributed to spin phonon coupling. We believe that our results will motivate the research community for further investigations towards understanding the material properties as well as in selecting material for appropriate device applications.

# ACKNOWLEDGEMENT

Authors acknowledge IISER Bhopal for research facilities, B. P acknowledges university grant commission for fellowship and S. S acknowledges DST/SERB for research funding (SERB project No. ECR/2016/001376 and Nano-mission project No. SR/NM/NS-84/2016(C)).